\documentclass[10pt,superscriptaddress,twocolumn,amsmath,amssymb,aps,prl,showpacs]{revtex4-1}
\usepackage{mathrsfs}
\usepackage{graphicx}
\usepackage{dcolumn}
\usepackage{bm}

\usepackage{graphicx}
\usepackage{epstopdf}

\usepackage{amsmath,amssymb,mathrsfs}
\usepackage{braket}
\usepackage{hyperref}
\hypersetup{
    colorlinks=true,
    linkcolor=blue,
    filecolor=blue,
    urlcolor=blue,
    citecolor=magenta
    }

\begin{document}


\title{The Role of the Effective Range in Resonantly Interacting Fermi Gases: \\ How Breaking Scale Symmetry Affects the Bulk Viscosity}


\author{Jeff Maki}
\affiliation{Department of Physics and HKU-UCAS Joint Institute for Theoretical and Computational Physics at Hong Kong, The University of Hong Kong, Hong Kong, China}
\author{Shizhong Zhang}
\email[]{shizhong@hku.hk}
\affiliation{Department of Physics and HKU-UCAS Joint Institute for Theoretical and Computational Physics at Hong Kong, The University of Hong Kong, Hong Kong, China}


\date{\today}

\begin{abstract} 
We investigate the role of the effective range on the bulk viscosity of $s$- and $p$-wave Fermi gases. At resonance, the presence of the effective range breaks the scale invariance of the system, and hence results in a non-zero bulk viscosity. However, we show that the effective range plays a very different role in the two cases. In the $s$-wave case, the role of the effective range is perturbative, and its contribution to the bulk viscosity vanishes in the limit of zero effective range. On the other hand, the effective range in $p$-wave Fermi gases leads to a non-zero bulk viscosity, even in the zero-range limit. We setup the general diagrammatic approach to compute the bulk viscosity spectral function that includes the effects of the effective range.  We then compute the analytic expressions for the spectral function in the high temperature limit, at low- and high-frequencies. We also derive the sume rules for the bulk viscosity spectral function for both $s$- and $p$-wave gases.
\end{abstract}
\maketitle

{\it Introduction}-- One of the most unique features of ultra-cold atomic gases is their universality~\cite{BECBCS_BOOK}. At low energies, the details of the short-range inter-atomic potential are irrelevant, and only a few low-energy scattering parameters are needed to describe the low-energy properties of the system~\cite{Pethick2008,LevSandro}. In the case of $s$-wave interactions, it is known that the scattering length, $a_0$, is the only relevant low-energy parameter for the so-called broad Feshbach resonance~\cite{Chin10}. At resonance, when $a_0=\pm\infty$, the system acquires scale symmetry, which imposes severe constraints on the thermodynamic~\cite{Ho04, Thomas05}, and transport properties of the system, the most amazing of which is the vanishing of the bulk viscosity~\cite{Son07, Nishida07, Enss2011,Taylor12}.

In real systems, however, a non-zero effective range always exists which breaks the scale symmetry of the system at resonance. The thermodynamic and transport properties of the system will then  depend on the effective range. To be specific, let $\delta_\ell(k)$ be the phase shift for two colliding particles with relative wave number, $k$, and angular momentum, $\ell$, then~\cite{LandauQM}:
\begin{align}
k^{2\ell+1}\cot\delta_\ell(k)=-\frac{1}{a_\ell}-R_\ell k^2+\cdots
\label{eq:phase_shift}
\end{align}
We will refer to $a_\ell$ as the scattering length, and $R_\ell$ as the effective range, noting that for $p$-wave scattering, $a_1$ has dimension of volume and $R_1$ has dimension of wave number. In actual experiments, a Feshbach resonance allows one to tune the scattering length, $a_{\ell}$, while the effective range, $R_{\ell}$ is approximately a constant. For any given system, the effective range, $R_{\ell}$, will be determined by the range of the inter-atomic interaction, $r_0$. As a result, Eq.~(\ref{eq:phase_shift}) is only valid at low-energies $k\ll r_0^{-1}$. For $s$-wave, this means that $k\ll R_0^{-1}$ while for $p$-wave, $k \ll R_1$.

To see the effects of the effective range, let us consider, for example, the binding energy of the two-body bound state, $E_b$, close to threshold, when $R_\ell$ is small. In the $s$-wave case, $E_b=-(\hbar^2/m)(1/a_0^2-2R_0/a_0^3)$ and tends to the universal result $-\hbar^2/(ma_0^2)$ when $R_0\to 0$; the effect of $R_0$ is perturbative. On the other hand, for the $p$-wave case~\cite{Bertulani2002, Braaten2012}, the existence of the shallow bound state depends crucially on the effective range: $E_b=-\hbar^2/(ma_1R_1)$. It is thus impossible to set $R_1=0$ while maintaining a finite bound state energy.  The role of the effective range on the energetic properties of the many-body system have been discussed for both $s$-wave~\cite{Pethick05,Forbes12, Lacroix16, Schonenberg2017,Miller18,Wellenhofer2020}, and $p$-wave cases~\cite{Gurarie2007, Yip05, Yip06, Zhang18, Zhang19}. 

In this Letter, we discuss the role of the effective range in the transport properties of both $s$- and $p$-wave Fermi gases, and compare the role of the effective range in these two cases. To this end, we calculate the bulk viscosity spectral function in the high temperature limit, and investigate how it depends on the effective range, particularly at resonance, {\it i.e.} when the scattering length is infinite: $a_\ell^{-1}  = 0$. In this limit, the viscosity spectral functions does not depend on $a_\ell$, and the role of the effective range is most transparent. If the effective range is an irrelevant quantity in the low-energy limit, scale invariance will be restored when one sets $R_\ell$ to zero. As a result, the bulk viscosity will necessarily vanish. However, if the effective range is relevant in the low-energy limit, it will be impossible to truly set the effective range to zero; the scale symmetry will then remain broken, and there will be a finite bulk viscosity.

{\it General setup}-- To describe both the resonant $s$- and $p$-wave Fermi gases in a unified scheme, we use the following two-channel Hamiltonian describing a system of fermions with spin $\sigma$, and $2\ell+1$ molecules interacting via the $\ell$-th partial wave channel (we set  $m=\hbar=1$):
\begin{align}
\widehat{H}_\ell &= \sum_{{\bf k}} \left(\frac{k^2}{2}-\mu\right) \psi_{\sigma}^{\dagger}({\bf k}) \psi_{\sigma}({\bf k}) \nonumber \\
&+ \sum_{{\bf Q}} \sum_{m=-\ell}^\ell \left( \frac{Q^2}{4} + \nu-2\mu \right)d^{\dagger}_m({\bf Q}) d_m({\bf Q}) \nonumber \\
&+g\sum_{m=-\ell}^\ell \sum_{{\bf Q},{\bf k}} \sqrt{\frac{4\pi}{V}} k^\ell \nonumber \\
&\left[Y_{\ell,m}({\bf \hat{ k}}) d^{\dagger}_m({\bf Q}) \psi_{\sigma}\left( \frac{{\bf Q}}{2}+{\bf k} \right) \psi_{\bar{\sigma}} \left( \frac{{\bf Q}}{2}-{\bf k} \right) + h.c. \right].
\label{eq:l_wave_H}
\end{align}
$\psi_{\sigma}({\bf k})$ is the field operator for the fermions with spin $\sigma$ while $d_m({\bf Q})$ is the field operator for the molecules with an azimuthal quantum number, $m$. $V$ is the volume of the system, $\nu$ is the detuning of the molecular field, and $g$ is the fermion-molecular coupling. We also let $\mu$ be the chemical potential, and $Y_{\ell,m}({\bf \hat{k}})$ is simply the $(\ell,m)$-spherical harmonic function. It is important to note that for $s$-wave interactions we assume the fermions are spinful: $\sigma = -\bar{\sigma}=1/2$, while they are spinless for the $p$-wave interactions $\sigma = \bar{\sigma}$. Here we have assumed that the scattering potential is isotropic in space, and the coupling constants $\nu$ and $g$ are independent of the azimuthal quantum number, $m$. By examining the two-body scattering of this model and comparing it to Eq.~(\ref{eq:phase_shift}), the bare coupling constants $\nu$ and $g$ can be related to the low-energy scattering parameters in the effective range expansion. For the $s$-wave case, we have $1/(4\pi a_0) = {\Lambda}/(2\pi^2) - {\nu}/{g^2}$ and   $R_0 = {4\pi}/{g^2}$, while for $p$-wave interactions: ${1}/(2\pi a_1) = {\Lambda^3}/({3\pi^2}) - {\nu}/{g^2}$ and ${R_1}/{2\pi} ={\Lambda}/\pi^2 + {1}/{g^2}$, where $\Lambda$ is an ultraviolet cutoff. The two-channel formulation has been successful in describing various thermodynamic properties of both $s$-wave~\cite{BECBCS_BOOK,Pethick05} and $p$-wave gases~\cite{Zhang19,Ohashi17,Gurarie2007}.

{\it Bulk viscosity with finite effective range}-- With the theory renormalized, we can now proceed to investigate the bulk viscosity for both the $s$- and $p$-wave gas. Using Kubo's formalism, the bulk viscosity is defined via the retarded correlation function of the stress-energy tensor~\cite{Nishida07, Taylor12}:
\begin{align}
\zeta(\omega) &= \frac{{\rm Im}\left[ \chi^{R}(\omega)\right]}{9\omega}, \label{eq:bulk_visc}\\
\chi^{R}(\omega) &= i\int_0^{\infty} dt \ e^{-i \omega t} \left\langle \left[\widehat{\Pi}(t), \widehat{\Pi}(0)\right] \right\rangle,
\label{eq:bulk_defs}
\end{align}
where the brackets denote the thermal average at temperature $T=1/\beta$, and $\langle \widehat{\Pi}(t)\rangle= 3 p(t) V$ is the trace of the stress-energy tensor at time $t$, and $p(t)$ is the pressure operator. As discussed in a recent work~\cite{Fujii2020}, this formula, as it stands, requires modification since it neglects the contributions to the bulk viscosity from pressure fluctuations. However, in the regimes of interest to us here, we obtain the same results using Eqs.~(\ref{eq:bulk_visc}, \ref{eq:bulk_defs}). By employing a scaling analysis, one can show that \cite{Pressure_Relation}:
\begin{equation}
\widehat{\Pi}(t) = 2 \widehat{H}_\ell + \frac{(2\ell+1) \widehat{C}_{a_\ell}(t)}{a_\ell} + (2\ell-1) \widehat{C}_{R_\ell}(t)R_\ell.
\label{eq:pressure_relation}
\end{equation}
Here we have defined two thermodynamic contact operators as the derivatives of the Hamiltonian with respect to the scattering length and the effective range: $\widehat{C}_{a_\ell}=-{\partial \widehat{H}}/{\partial a_{\ell}^{-1}}$ and $\widehat{C}_{R_\ell}(t) =-{\partial \widehat{H}}/{\partial R_{\ell}}$. Explicitly, they are given by:
\begin{align}
\widehat{C}_{a_\ell}(t) &=\sum_{m=-\ell}^{\ell} \frac{g^2}{4\pi A_\ell}\int d{\bf R}d_m^{\dagger}({\bf R},t)d_m({\bf R},t) \\
\widehat{C}_{R_\ell}(t) &=\sum_{m=-\ell}^{\ell} \frac{g^2}{4\pi A_\ell}\int d{\bf R}d_m^{\dagger}({\bf R},t)\left(i \partial_t + \frac{\nabla^2}{4}\right)d_m({\bf R},t),
\label{eq:def_contacts}
\end{align}
where $A_\ell = 1- \delta_{\ell,1} /2$, is a symmetry factor that represents the indistinguishable nature of the spinless fermions considered when $\ell=1$. The thermodynamic contacts represent the change in the energy with respect to the scattering length, $a_\ell$, and the effective range, $R_\ell$, respectively~\cite{Yoshida2015, Yu16,He2016,Peng2016}. We note that in order to obtain the current form of $\widehat{C}_{R_\ell}$, we have integrated out the fermionic degrees of freedom using the Heisenberg equation of motion for the molecular field.

\begin{figure}
\includegraphics[scale=0.4]{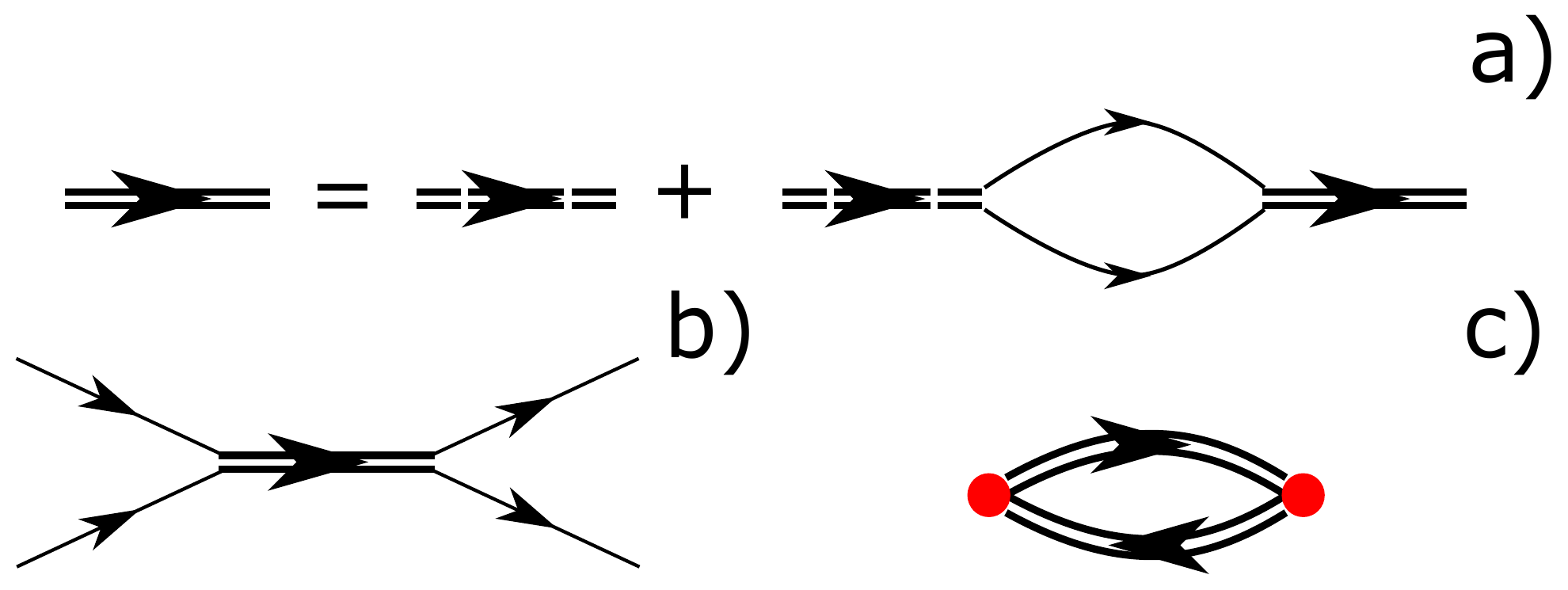}
\caption{Diagrammatic calculation for contact-contact correlation functions. (a) The full molecular propagator (solid double line) is given by the bare molecular propagator (dashed double line) dressed by fermionic pair propagation (single solid line). (b) The diagram for the scattering amplitude of two fermions. (c) The diagram for the calculation of bulk viscosity. Each vertex (red) represents an insertion of a contact operator $\widehat{C}_{a_\ell}$ or $\widehat{C}_{R_\ell}$, see Eqs.~(\ref{eq:correlator_AA}-\ref{eq:correlator_RR}).}
\label{fig:NRS_Scheme}
\end{figure}

After substituting Eq.~(\ref{eq:pressure_relation}) into Eq.~(\ref{eq:bulk_defs}), one can see that the calculation of the bulk viscosity reduces to the evaluation of three retarded correlators~\cite{ENERGY_NOTE}:
\begin{align}
\chi^R_{a,a}(\omega) &= i\frac{(2\ell+1)^2}{a_\ell^2}\int_0^{\infty} dt e^{-i \omega t}\langle [\widehat{C}_{a_\ell}(t), \widehat{C}_a(0)] \rangle \label{eq:correlator_AA}\\
\chi^R_{a,R}(\omega) &=i\frac{(2\ell-1)(2\ell+1)R_\ell}{a_\ell}\int_0^{\infty} dt e^{-i \omega t} \langle [\widehat{C}_{a_\ell}(t), \widehat{C}_{R_\ell}(0)] \rangle \label{eq:correlator_AR} \\
\chi_{R,R}^R(\omega) &= i (2\ell-1)^2R_\ell^2 \int_0^{\infty} dt e^{-i \omega t}\langle [\widehat{C}_{R_\ell}(t), \widehat{C}_{R_\ell}(0)]\rangle.
\label{eq:correlator_RR}
\end{align}
It is possible to evaluate the correlation functions Eqs.~(\ref{eq:correlator_AA}-\ref{eq:correlator_RR}) diagrammatically as depicted in Fig.~\ref{fig:NRS_Scheme}. Here the molecular propagator is dressed by the process of dissociation and association of two fermions. In Matsubara space, followed by the standard analytic continuation, one can then show that the bulk viscosity in the normal phase is explicitly given by $\zeta_\ell(\omega)  = \zeta_{a,a}(\omega)+2 \zeta_{a,R}(\omega) + \zeta_{R,R}(\omega)$ with:
\begin{widetext}
\begin{align}
\zeta_{a,a}(\omega) &= \frac{(2\ell+1)^2}{9a_\ell^2 V}\sum_{m,{\bf Q}} \int_{-\infty}^{\infty} \frac{dx}{\pi} {\rm Im} \left[ \frac{g^2D({\bf Q}, x+i \delta)}{4\pi A_\ell}\right]{\rm Im} \left[ \frac{g^2D({\bf Q}, x+\omega +  i \delta)}{4\pi A_\ell}\right]f(x,\omega) \label{zetaaa}\\
\zeta_{a,R}(\omega) &= \frac{(2\ell-1)(2\ell+1)R_\ell}{9 a_\ell V} \sum_{m,{\bf Q}} \int_{-\infty}^{\infty} \frac{dx}{\pi} {\rm Im} \left[ \frac{g^2D({\bf Q}, x+i \delta)}{4 \pi A_l}\right]{\rm Im} \left[ \frac{g^2D({\bf Q}, x+\omega +  i \delta)}{4 \pi A_l}\right] f(x,\omega)\left(x-\frac{Q^2}{4} + 2\mu + \frac{\omega}{2}\right) \label{zetaaR}\\
\zeta_{R,R}(\omega) &= \frac{(2\ell-1)^2R_\ell^2}{9 V}\sum_{m,{\bf Q}} \int_{-\infty}^{\infty} \frac{dx}{\pi}{\rm Im} \left[\frac{g^2D({\bf Q}, x + i \delta)}{4\pi A_\ell}\right]{\rm Im} \left[\frac{g^2D({\bf Q}, x+\omega + i \delta)}{4\pi A_\ell}\right] f(x,\omega)\left(x-\frac{Q^2}{4} + 2\mu + \frac{\omega}{2}\right)^2,
\label{zetaRR}
\end{align}
\end{widetext}
where $f(x,\omega)=\omega^{-1}[n_B(x) - n_B(x+\omega)]$ and $n_B(x) = [\exp(\beta x) -1]^{-1}$ is the Bose-Einstein distribution function. $D({\bf Q},x)$ is the molecular propagator with momentum ${\bf Q}$ and frequency $x$, evaluated in the presence of a thermal fermionic background~\cite{Zhang18}. Eqs.~(\ref{zetaaa}-\ref{zetaRR}) provide the general expressions for the bulk viscosity for both $s$- and $p$-wave Fermi gases while including the effective range \cite{INTEGRATING_NOTE}.

In the high temperature limit, it is possible to neglect the presence of the thermal fermionic background, and use the two-body results. In this limit, Eq.~(\ref{zetaaa}) for $s$-wave interactions is consistent with the results obtained in Refs.~\cite{Enss19, Nishida19, Hofmann20}. A similar expression was also obtained in~\cite{Maki19} for one-dimensional Fermi gases with three-body $s$-wave interactions. In addition to this term, the presence of the effective range has generated two additional contributions to the bulk viscosity. We now examine how these additional terms modify the bulk viscosity spectral function.

At high frequencies, the bulk viscosity spectral function exhibits power law tails, while the magnitude of the tail is proportional to the contacts. In practice, the high frequency limit is determined by setting the frequency much larger than the bound state energy: $\omega\gg a_0^{-2}$ for $s$-wave and $\omega\gg 1/(a_1 R_1)$ for $p$-wave, but much smaller than the scale set by the effective range: $\omega \ll  R_0^{-2}$ for $s$-wave interactions and $\omega \ll R_1^{2}$ for $p$-wave interactions. For frequencies larger than the scale set by the effective range, our effective field theory is inapplicable, as mentioned previously. Evaluating Eqs.~(\ref{zetaaa}-\ref{zetaRR}) at high-frequencies, and in the high temperature limit, yields the following divergent terms:
\begin{align}
\zeta^d_0(\omega)&\approx \frac{\sqrt{\omega}R_0^2}{36 V}\left[C_{a_0} \left(1-\frac{1}{a_0^2 \omega}-\frac{4}{a_0 R_0 \omega}\right)+\frac{C_{R_0}}{\omega} \right],\label{eq:high_freq_limit-s} \\
\zeta^d_1(\omega) &\approx \frac{\sqrt{\omega}}{36V}\left[C_{a_1}\left(1+\frac{10}{a_1 R_1 \omega}\right) + \frac{7C_{R_1}}{2\omega} \right],\label{eq:high_freq_limit-p}
\end{align}
for $s$-wave ($\zeta^d_0$), and $p$-wave ($\zeta^d_1$) respectively. Eqs.~(\ref{eq:high_freq_limit-s}, \ref{eq:high_freq_limit-p}) represent the divergent high-frequency terms that enter the viscosity sum rule (discussed below), and arise solely from the contributions due to the effective range, $\zeta_{R,R}(\omega)$ and $\zeta_{a,R}(\omega)$. This is one of the main results of this work. 

Upon examining Eqs.~(\ref{eq:high_freq_limit-s}, \ref{eq:high_freq_limit-p}), it is possible to understand the relevancy of the effective range in the $s$- and $p$-wave cases. In the case of $s$-wave interactions, it is possible to set the effective range $R_0\to 0$ while still being consistent with the requirement: $\omega\ll R_0^{-2}$. Since the divergent term $\zeta_0^d(\omega)$ is proportional to $R_0^2$, the high frequency tails simply vanish when $R_0\to 0$. From Eq.~(\ref{zetaaa}), one reproduces the standard result that:
\begin{equation}
\zeta_0(\omega, R_0 \rightarrow 0) = \frac{1}{9 a^2} \frac{1}{\omega^{3/2}} \frac{C_a}{V}.
\end{equation}

This expression is equivalent to the single channel model results obtained previously~\cite{Taylor12, Nishida19, Enss19, Hofmann20}. This result reinforces the common procedure of taking the zero-range limit for $s$-wave interactions. In the limit of a small, but finite, effective range, $R_0$, the correction to the zero-range bulk viscosity is perturbative in nature, and can be ignored. On the other hand, the role of the effective range, $R_0$, is most apparent at unitarity, when $a_0\to\pm\infty$. In this limit, the presence of the effective range breaks the scale invariance of the unitary Fermi gas, and leads to a finite bulk viscosity. The bulk viscosity will entirely be due to $\zeta_{R,R}(\omega)$ in Eq.~(\ref{zetaRR}), and yields the following divergent term:
\begin{equation}
\zeta^d_0(\omega, a_0 \rightarrow \pm \infty) = \frac{R_0^2C_{a_0}}{36 V} \sqrt{\omega}.
\end{equation}
At resonance, the leading term in the contact $C_{a_0}$ is a constant, independent of $R_0$, and therefore $\zeta_0^d(\omega, a_0 \to \pm \infty) \propto R_0^2$.

In the case of $p$-wave interactions, it is clear that we cannot take the zero-range limit, $R_1\to 0$, while maintaining the high frequency regime: $1/(a_1 R_1)\ll \omega\ll R_1^2$. In practice, $R_1$ is finite, and a well-defined high frequency limit can always be achieved close to resonance. For instance, $^{40}$K has $R_1\approx 25 k_F$, which yields a low-energy window: $0\ll \omega\ll 600\epsilon_F$. In this regime, the high frequency tails of $\zeta_1(\omega)$ will certainly be relevant in actual experiments. At high-frequencies, one finds the same $\sqrt{\omega}$-dependence, as in the $s$-wave case, with a coefficient proportional to $C_{a_1}$:
\begin{equation}
\zeta^d_1(\omega, a_1 \rightarrow \pm\infty) =\frac{C_{a_1}}{36 V} \sqrt{\omega},
\end{equation}
where the contact $C_{a_1} \propto R_1^{-1}$, and so $\zeta^d_1(\omega, a_1 \to \pm\infty) \propto R_1^{-1}$. This expression is divergent in the {\it zero-range} limit, when $R_1 \to 0$, in contradistinction to the $s$-wave case. As in the $s$-wave case, the presence of the $p$-wave effective range breaks the scaling symmetry of the system, and results in a non-zero bulk viscosity. However, its effect is dominant in the zero-range limit for the $p$-wave case.

The importance of the effective range can also be seen in the low-frequency limit. Here we report the bulk viscosities strictly at resonance when $a_\ell^{-1} =0$. The relevant high temperature limit for the $s$-wave case is given by: $1/a_0^2\ll k_BT \ll R_0^{-2}$, and for $p$-wave case: $1/(a_1 R_1)\ll k_B T\ll R_1^2$. One can again see that in the $s$-wave case, the high temperature limit can be safely taken in the zero range limit, while it is impossible for the $p$-wave case. From Eqs.~(\ref{zetaaa}-\ref{zetaRR}), one obtains:
\begin{align}
\zeta_0(\omega \rightarrow 0) &\approx \frac{2^{5/2}z^2}{9} \frac{R_0^2}{\lambda_T^5} \propto T^{5/2}R_0^2\label{eq:zero_frequency-s}, \\
\zeta_1(\omega \rightarrow 0) &\approx \frac{2^{5/2}z^2}{3} \frac{1}{R_1^2\lambda_T^5} \propto \frac{T^{5/2}}{R_1^2}\label{eq:zero_frequency-p},
\end{align} 
where $z = \exp(\beta \mu)$ is the fugacity, and $\lambda_T = \sqrt{2\pi \beta}$ is the thermal wave-length  \cite{NONANALYTICALTERMS}. Although Eqs.~(\ref{eq:zero_frequency-s}, \ref{eq:zero_frequency-p}) look similar, we note that due to the physical requirement of low energies, the effective range have opposite roles in the $s$- and $p$-wave case. When one takes the zero range limit, the bulk viscosity is vanishingly small for $s$-wave gases, but diverges for $p$-wave gases.

{\it Sum rules for bulk viscosity spectral function}.-- Finally, we report the sum rules for the bulk viscosity spectral function. As noted from Eqs.~(\ref{eq:high_freq_limit-s}, \ref{eq:high_freq_limit-p}), the spectral function is divergent. In order to have a well defined sum rule, it is necessary to remove these divergent pieces, which we have labeled as: $\zeta_{\ell}^{d}(\omega)$. Once that is done, it is possible to use the Kramers-Kronig relations to show that $\zeta_\ell(\omega)$ satisfies the following sum rules:
\begin{align}
&\int_{-\infty}^{\infty} \frac{d\omega}{\pi} \left(\zeta_\ell(\omega) - \zeta_\ell^d(\omega)\right) \nonumber \\
&= - \frac{1}{9V}\left[(2\ell+1)^2 \frac{\partial C_{a_\ell}}{\partial a_\ell} + \frac{\partial C_{R_\ell}}{\partial R_\ell^{-1}} \right. \nonumber \\
&\left. +(2\ell-1)(2\ell+1) \left( \frac{1}{a_\ell R_\ell} \frac{\partial C_{a_\ell}}{\partial R_\ell^{-1}} + a_\ell R_\ell \frac{\partial C_{R_\ell}}{\partial a_\ell}\right)\right].
\label{eq:sum_rule}
\end{align}
This relation generalizes the already well-known sum rule for the $s$-wave Fermi gases derived in Ref.~\cite{Taylor12, Nishida19, Enss19, Hofmann20} to include the presence of the effective range, as well as being applicable for $p$-wave gases.

{\it Conclusions}-- In this work we illustrated how the bulk viscosity spectral function depends on the effective range for both $s$- and $p$-wave Fermi gases. We focused on the case of resonantly interacting Fermi gases in the high-temperature limit, where the effects of the effective-range are most transparent.  We show that the presence of the effective range introduces a second thermodynamic contact, and as a result, new terms to the bulk viscosity. For $s$-wave Fermi gases, the effective range is an irrelevant parameter; the low-energy and zero-range limits can be simultaneously taken. The bulk viscosity will then vanish alongside the effective range at resonance. While for $p$-wave gases the situation is different, the effective range is a relevant parameter; the low-energy and zero-range limits can not be simultaneously taken. Therefore, even at resonance, the presence of the effective range will lead to a finite bulk viscosity. This work clearly reinforces the traditional prescription of taking the zero-range limit for $s$-wave gases, for both the energetics and transport dynamics, and shows how the $p$-wave gas is fundamentally different.

In principle, one could extend this calculation to gases interacting via higher partial waves, and to arbitrary order in the effective range expansion. However, the calculation becomes increasingly complex, as one needs to consider the correlations between the different contact operators. A simple scaling analysis shows that for a given partial wave, $\ell$, the first $\ell +1$ terms in the effective range expansion are relevant; with the higher order terms irrelevant. This scaling analysis is consistent with the results obtained in this Letter.

\paragraph{Acknowledgements}
The authors would like to thank Yusuke Nishida for useful discussions. This work is supported by HK GRF 17318316, 17305218 and CRF C6026-16W and C6005-17G, and the Croucher Foundation under the Croucher Innovation Award.


\begin{thebibliography}{References}

\bibitem{BECBCS_BOOK} {\it The BCS-BEC Crossover and the Unitary Fermi Gas}, ed. Wilhelm Zwerger (Springer, 2011).

\bibitem{Pethick2008}
C. J. Pethick and H. Smith, \textit{Bose-Einstein Condensation in Dilute Gases}, 2nd edition, Cambridge University Press (2008).

\bibitem{LevSandro} Lev Pitaevskii and Sandro Stringari, \textit{Bose-Einstein Condensation and Superfluidity}, 2nd edition, Oxford University Press (2016). 



\bibitem{Chin10} C. Chin, R. Grimm, P. Julienne, and E. Tiesinga, Rev. Mod. Phys. {\bf 82}, 1225 (2010).

\bibitem{Ho04} T.-L. Ho, Phys. Rev. Lett. {\bf 92}, 090402 (2004).

\bibitem{Thomas05} J. E. Thomas, J. Kinast, and A. Turlapov, Phys. Rev. Lett. {\bf 95}, 120402 (2005).

\bibitem{Son07} D. T. Son, Phys. Rev. Lett. {\bf 98}, 020604 (2007).

\bibitem{Nishida07} Y. Nishida, D. T. Son, Phys. Rev. D {\bf 76}, 086004 (2007).

\bibitem{Enss2011} T. Enss, R. Haussmann, and W. Zwerger, Ann. Phys. {\bf 326}, 770 (2011)

\bibitem{Taylor12} E. Taylor, and M. Randeria, Phys. Rev. Lett. {\bf 109}, 135301 (2012).



\bibitem{LandauQM} L.D. Landau, and E. M. Lifshitz, {\it Quantum Mechanics: Non-Relativistic Theory}, (Pergamon Press, Toronto, Canada, 1991).

\bibitem{Bertulani2002} C.A.Bertulani,  H.-W.Hammer, U.van Kolck, Nuclear Physics A, {\bf 712}, 37 (2002)

\bibitem{Braaten2012} E. Braaten, P. Hagen, H.-W. Hammer, and L. Platter, 
Phys. Rev. A {\bf 86}, 012711 (2012).?


\bibitem{Pethick05} A. Schwenk, and C. J. Pethick, Phys. Rev. Lett. {\bf 95}, 160401 (2005).

\bibitem{Forbes12} M. M.N. Forbes, S, Gandolfi, and A. Gezerlis, Phys. Rev. A {\bf 86}, 053603 (2012).

\bibitem{Lacroix16} D. Lacroix, Phys. Rev. A {\bf 94}, 043614 (2016).

\bibitem{Schonenberg2017} L. M. Schonenberg and G. J. Conduit, Phys. Rev. A {\bf 95}, 013633 (2017).

\bibitem{Miller18} G. A. Miller, Phys. Lett. B {\bf 777}, 442 (2018).

\bibitem{Wellenhofer2020} C. Wellenhofer, C. Drischler and A. Schwenk, Physics Letters B {\bf 802}, 135247 (2020).



\bibitem{Yoshida2015} 
S. M. Yoshida and M.Ueda, Phys. Rev. Lett. \textbf{115}, 135303 (2015).


\bibitem{Yu16}
Z. Yu, J. H. Thywissen, and S. Zhang, Phys. Rev. Lett. {\bf 115}, 135304 (2015). Erratum, {\em ibid}. {\bf 117}, 019901 (2016).

\bibitem{He2016}
M. He, S. Zhang, H. M. Chan, and Q. Zhou, Phys. Rev. Lett. \textbf{116}, 045301 (2016).

\bibitem{Peng2016}
S.-G. Peng, X.-J. Liu, H. Hu, Phys. Rev. A {\bf 94} 063651 (2016).

\bibitem{Gurarie2007}
V. Gurarie, L. Radzihovshy, Ann. of Phys. {\bf 332}, 2-119 (2007).

\bibitem{Yip05} C.-H. Cheng and S.-K. Yip, Phys. Rev. Lett. {\bf 95}, 070404 (2005).
\bibitem{Yip06} C.-H. Cheng and S.-K. Yip, Phys. Rev. A {\bf 73}, 064517 (2006).

\bibitem{Zhang18} J. Yao, and S. Zhang, Phys. Rev. A {\bf 97}, 043612 (2018).

\bibitem{Zhang19} S. Ding, and S. Zhang, Phys. Rev. Lett. {\bf 123}, 070404 (2019).

\bibitem{Ohashi17} D. Inotani, P. van Wyk, and Y. Ohashi, J. Phys. Soc. Jpn. {\bf 86}, 024302 (2017).

\bibitem{Fujii2020} K. Fujii and Y. Nishida, arXiv:2004.12154v1 (2020)



\bibitem{Pressure_Relation} To obtain this result we note from dimensional analysis: $p(t) = \beta^{-5/2} G(\beta \mu, \beta^{(2\ell+1)/2} a_\ell^{-1}, \beta^{(2\ell-1)/2}R_\ell)$. Since $\exp \left(\beta p(t) V\right)  =  \mbox{Tr}[\exp(-\beta H)]$, one can obtain the pressure relation by considering the derivative of $p(t)$ with respect to $\beta$.


\bibitem{ENERGY_NOTE} The retarded correlators involving the Hamiltonian and a single contact are defined to vanish in thermal equilibrium, as it  is related to the time derivative of the contact. As described in \cite{Fujii2020}, this is appropriate if one considers finite frequencies.

\bibitem{INTEGRATING_NOTE} In our calculation, we neglect the interactions between the fermions and the dressed molecules. Such interactions can lead to a contribution to the bulk visocisty from four-body processes, {\it i.e.} the scattering of two molecules. We neglect these processes in this analysis.

\bibitem{Nishida19} Y. Nishida, Ann. of Phys. {\bf 410}, 167949 (2019).

\bibitem{Enss19} T. Enss, Phys. Rev. Lett. {\bf 123}, 205301 (2019).

\bibitem{Hofmann20} J. Hofmann, Phys. Rev. A {\bf 101}, 013620 (2020).

\bibitem{Maki19} J. Maki, and F. Zhou, Phys. Rev. A, {\bf 100}, 023601 (2019).

\bibitem{NONANALYTICALTERMS} We note that if one deviates from resonance the expression for $\zeta_{a,a}(\omega)$ is non-analytical when one considers the limits of resonance and zero-frequency. These two limits do not commute due to a logarithmic singularity. This issue is most relevant in the zero-range limit, where the dominant contribution to the bulk viscosity is from $\zeta_{a,a}(\omega)$. See Refs.~\cite{Nishida19, Enss19, Hofmann20}, and more recently~\cite{Fujii2020} for more details.


\end{thebibliography}
\end{document}